\documentclass[12pt]{iopart}
\begin{document}

\title{Wave functions and characteristic times for transmission and reflection}

\author{N L Chuprikov
\footnote[3]{Also at Physics Department, Tomsk State University} }

\address{Tomsk State Pedagogical University, 634041, Tomsk, Russia}

\begin{abstract}

We present a renewed wave-packet analysis based on the following ideas: if a quantum
one-particle scattering process and the corresponding state are described by an
indivisible wave packet to move as a whole at all stages of scattering, then they are
elementary; otherwise, they are combined; each combined process consists from several
alternative elementary ones to proceed simultaneously; the corresponding (normed)
state can be uniquely presented as the sum of elementary ones whose (constant) norms
give unit, in sum; Born's formula intended for calculating the {\it expectation}
values of physical observables, as well as the standard timing procedure are valid
only for elementary states and processes; only an elementary time-dependent state can
be considered as the quantum counterpart to some classical one-particle trajectory. By
our approach, tunneling a non-relativistic particle through a static one-dimensional
potential barrier is a combined process consisting from two elementary ones,
transmission and reflection. In the standard setting of the problem, we find an
unique pair of solutions to the Schr\"odinger equation, which describe separately
transmission and reflection. On this basis we introduce (exact and asymptotic)
characteristic times for transmission and reflection.

\end{abstract}


\maketitle

\newcommand{\Api}{A_{in}}
\newcommand{\Ami}{B_{in}}
\newcommand{\Apo}{A_{out}}
\newcommand{\Amo}{B_{out}}
\newcommand{\bpi}{a_{in}}
\newcommand{\bmi}{b_{in}}
\newcommand{\bpo}{a_{out}}
\newcommand{\bmo}{b_{out}}
\newcommand{\api}{a_{in}}
\newcommand{\ami}{b_{in}}
\newcommand{\apo}{a_{out}}
\newcommand{\amo}{b_{out}}

\section{Introduction}
For a long time tunneling a particle through an one-dimensional time-independent
potential barrier was considered in quantum mechanics as a representative of
well-understood phenomena. However, now it has been realized that this is not the
case. The inherent to quantum theory standard wave-packet analysis (SWPA)
\cite{Ha2,Wig,Har,Ha1,Ter} (see also \cite{Col}), in which the study of the temporal
aspects of tunneling is reduced to timing the motion of the center of "mass" (CM) of
the corresponding wave packet to describe the process, does not provide a clear
prescription both to interpret the scattering of finite in $x$ space wave packets and
to introduce characteristic times for a tunneling particle. The latter is known as
the tunneling time problem (TTP) which has been of great interest for the last
decades.

As is known (see \cite{La2}), the main difficulties to arise in interpreting the
wave-packet's tunneling are connected to the fact that there is no causal link between
the transmitted (or reflected) and incident wave packets. One of the visual
consequences of this is that the average particle's kinetic energy for the
transmitted, reflected and incident wave packets is different. For example, in the
case of an opaque rectangular barrier, the velocity of the CM of the transmitted wave
packet is larger than that of the incident one. It is evident that this fact needs a
proper physical explanation. As was pointed out in \cite{La2}, it would be strange to
interpret the above property of wave packets as the evidence of accelerating a
particle (in the asymptotic regions) by the static potential barrier.

One has to point also to the well-known Hartman effect \cite{Har} related to the
acceleration of the CM of the transmitted wave packet, to superluminal velocities
(see also \cite{Mu6}). In many respects, the present interpretation of this property
of wave packets is still controversial.

Note, in the case of wide (strictly speaking, infinite) in $x$ space wave packets the
average kinetic energy of particles, before and after the interaction, is the same.
However, it is evident that a causal link between the transmitted and incident wave
packets does not appear in this limiting case. Perhaps, this fact is a basic reason
by which many physicists appraise the phase times introduced in the SWPA as
ill-defined. At least, a review \cite{Ha2} devoted to the TTP seems to be the last one
in which the SWPA is considered in a positive context.

Apart from the SWPA to deal with the CM of a wave packet, in the same or different
setting the tunneling problem, a variety of alternative approaches (see reviews
\cite{Ha2,La1,Olk,Ste,Mu0,Nu0} and references therein) to introduce various
characteristic times for a tunneling particle have also been developed. Among the
alternative concepts, of interest are that of the dwell time \cite{Smi,Ja1,Le1,Nus},
that of the Larmor time \cite{Baz,Ryb,But,Bu1,Zhi} to give the way of measuring the
dwell time, and the concept of the time of arrival which is based on introducing
either a suitable time operator (see, for example, \cite{Aha,Mu4,Hah,Noh,Mu9}) or the
positive operator valued measure (see review \cite{Mu0}). Besides, of importance are
the studies of the temporal aspects of tunneling on the basis of the Feynman, Bohmian
and Wigner approaches to deal with the random trajectories of particles (see, for
example, \cite{Sok,Bo1,Yam,Ymm,Gru} and references therein). One should also mention
the papers \cite{Ga1,Ga2,Ga3} where the TTP is studied beyond the framework of the
standard setting the scattering problem.

We have to stress however that in the standard setting, when the initial wave packet
may include a zero-momentum component, none of the alternative approaches have led to
commonly accepted characteristic times (see \cite{Ha2,La1,Olk,Ste,Mu0,Nu0}). The
recent papers \cite{Got,Kwe}, which present new versions of the dwell time (see
\cite{Got}) and complex tunneling times (see \cite{Kwe}), evidence too that up to now
there are no preferable time concepts for a tunneling particle.

There is an opinion that the TTP, in the standard setting of the one-dimensional
scattering problem, is ill-defined, since it does not include the measurement process.
We think that such an opinion is very questionable. Of course, in some cases the
measurement process can essentially modify the original scattering one, and hence the
study of its possible influence on the temporal aspects of this scattering process can
be very useful (see, for instance, \cite{Mu3} and the review \cite{Mu0} where this
question is deeply analyzed). At the same time, to state that any
measurement-independent setting of the TTP is ill-defined is unacceptable, in
principle. Otherwise, all Hamiltonians to describe a measurement-independent
scattering processes (including the motion of a free particle, and the tunneling
process) would be considered as ones having no physical sense. We have to stress that
any quantum scattering process, like classical one, proceeds, irrespective of our
assistance, in some space-time framework. So that quantum theory should give a clear
and unambiguous prescription to define both the spatial and temporal limits of this
process.

The main question of the TTP, which implies an unique answer, is that of the (average)
time spent by a quantum particle inside a finite barrier region. There is a
particular case when answering this question is trivial. We have in mind tunneling a
particle through the $\delta$-potential barrier. Indeed, one can {\it a priori} say
that this characteristic time should be equal to zero for this potential. For the
probability to find a particle in its barrier region is equal to zero.

Note, the TTP is very often (see, for example, \cite{Ol1}) treated as the problem of
introducing characteristic times for {\it a wave packet} passing through a
quantum-mechanical barrier. As is seen, from the very outset, this formulation implies
timing a lengthy object whose spatial size is comparable with (or even much more
than) the width of the potential barrier. Such a vision of the TTP is of wide
spreading. Therefore it is no mere chance that a nonzero phase transmission time
obtained in the SWPA for the $\delta$-potential is viewed by many physicists as a
fully expected result, which allegedly says about the non-locality of a quantum
scattering process. However, this result, being derived in the SWPA on the basis of
timing the motion of the CM of a wave packet, is {\it a priori} inconsistent: any
selected point of a wave packet should cross instantaneously the point-like support
of the $\delta$-potential.

As regards the non-locality of tunneling, the example of the $\delta$-potential shows
explicitly that the time spent by a quantum particle in the barrier region provides
insufficient information about a quantum one-particle scattering process. It is
useful also to define the time interval when the probability to find a particle
crossing through the barrier region is sufficiently large. The necessity in the
additional time scale is associated eventually with the fact that the time of arrival
of a particle at some point can be predicted, in quantum theory, with the error
amounting to the half-width of the corresponding wave packet. It is this
characteristic time that must be derived, for a particle, with taking into account of
the wave-packet's width. This time, which can be treated as the time of the
interaction of a quantum ensemble of particles with the barrier, is always greater
than the time spent in the barrier region by each particle of this ensemble. It is
this quantity that must be nonzero for the $\delta$-potential.

So, due to the uncertainty in finding the position of a tunneling particle, the
(average) time spent by the particle in the barrier region is insufficient to give a
full information about its interaction with the barrier. However, there is once more
peculiarity of a quantum description of the one-dimensional scattering of a particle,
which drastically complicates solving the TTP. Indeed, in classical theory, in timing
a scattering particle for a given initial condition, we deal with the only trajectory
of a particle, that corresponds either transmission or reflection. However, in
quantum description we deal with a wave function to include information about both
the alternative possibilities. Therefore every physicist setting to the TTP has
firstly to resolve the dilemma, whether he has to introduce individual (transmission
and reflection) times or whether he must solve the TTP with no distinguishing between
transmission and reflection.

One should recognize that at present this question is still open. Most of the time
concepts, such as the time of arrival as well as the dwell, Larmor and phase
tunneling times suggest introducing individual characteristic times for transmission
and reflection. As is pointed out by Nussenzveig (see \cite{Nu0}), "\dots [if some
characteristic time] does not distinguish between reflected and transmitted
particles, [this is] usually taken as a defect \dots". At the same time, Nussenzveig
himself believes (ibid) that "\dots [a joint description of the whole ensemble of
tunneling particles] is actually a virtue, since transmission and reflection are
inextricably intertwined; \dots only the characteristic times averaged over
transmitted and reflected particles have a physical sense".

An intrigue is that there are forcible arguments in both the cases. On the one hand,
quantum mechanics, as it stands, indeed provides no prescription to separate
to-be-transmitted and to-be-reflected particles at the early stages of scattering.
Thus, having no information about the behaviour of both the kinds of particles in the
barrier region, it is impossible to find the average time spent, in this region, by
particles of each kind. A knowledge about their behavior after the scattering event is
insufficient for this purpose.

On the other hand, the final state of a tunneling particle evidences that tunneling
consists in fact from two alternative processes - transmission and reflection. Born's
formula underlying the statistical interpretation of quantum mechanics fails in this
case: the average values of the particle's position and momentum calculated over the
whole ensemble of particles, cannot be interpreted as the expectation values of these
quantities. We consider that this fact is a poor background for introducing
characteristic times averaged over transmitted an reflected particles.

In fact, the above controversy says that usual quantum mechanics does not provide
both a joint and separate description of transmitted and reflected particles. It
enables one to study in detail the temporal behavior of wave packets to describe the
tunneling process. However, it gives no basis to extract from these detailed data the
expectation values of the particle's position and momentum, as well as to introduce
its characteristic times. Its basic tools - Born's formula for calculating the
expectation values, and the standard timing procedure - proved to be usefulness in
studying a tunneling particle.

The main idea of this paper is that in order to learn to calculate expectation values
of physical observables for a tunneling particle (and solve the TTP, on this basis)
one needs to correct our understanding of the nature of a quantum one-particle
scattering state and the correspondence principle. Now it is generally accepted that
any quantum time-dependent one-particle state can be considered, in principle, as the
quantum counterpart to some classical one-particle trajectory. However, generally
speaking, this is not the case.

In this approach, all quantum one-particle scattering processes described by the
Schr\"odinger equation are divided into two classes - combined and elementary. If the
wave packet to describe a quantum one-particle scattering process represents at some
time a disconnected object (or, in other words, when the set of spatial points, where
the probability to find a particle is nonzero, is disconnected), then we deal with a
combined process. Otherwise, the process is elementary. Only in the last case, Born's
formula and the standard timing procedure are applicable. By our approach, only an
elementary time-dependent one-particle scattering state can be considered as the
quantum counterpart to some classical one-particle trajectory. As regards a combined
time-dependent state, it can be associated with several one-particle trajectories.

On the basis of this idea we develop a renewed wave-packet analysis in which we treat
the one-particle one-dimensional scattering of a particle on a static potential
barrier as a combined process consisting from two alternative ones, transmission and
reflection. We hope that this approach will be useful for a deeper understanding of
the nature of quantum one-particle scattering processes and, in particular, the
tunneling effect.

The paper is organized as follows. In Section \ref{a1} we pose a complete
one-dimensional scattering problem for a particle. Shortcomings of the SWPA are
analyzed in Section \ref{a12}. In Section \ref{a2} we present a renewed wave-packet
analysis in which transmission and reflection are treated separately. In Section
\ref{a3} we define the average (exact and asymptotic) transmission and reflection
times and consider the cases of rectangular barriers and $\delta$-potentials. In the
last section some aspects of our approach are discussed in detail.

\newcommand{\ko}{\kappa_0^2}
\newcommand{\kj}{\kappa_j^2}
\newcommand{\kd}{\kappa_j d_j}
\newcommand{\kki}{\kappa_0\kappa_j}

\newcommand{\Ra}{R_{j+1}}
\newcommand{\Rb}{R_{(1,j)}}
\newcommand{\Rc}{R_{(1,j+1)}}

\newcommand{\Ta}{T_{j+1}}
\newcommand{\Tb}{T_{(1,j)}}
\newcommand{\Tc}{T_{(1,j+1)}}

\newcommand{\Wa}{w_{j+1}}
\newcommand{\Wb}{w_{(1,j)}}
\newcommand{\Wc}{w_{(1,j+1)}}

\newcommand{\UU}{u^{(+)}_{(1,j)}}
\newcommand{\VV}{u^{(-)}_{(1,j)}}

\newcommand{\ta}{t_{j+1}}
\newcommand{\tb}{t_{(1,j)}}
\newcommand{\tc}{t_{(1,j+1)}}

\newcommand{\tee}{\vartheta_{(1,j)}}

\newcommand{\tta}{\tau_{j+1}}
\newcommand{\ttb}{\tau_{(1,j)}}
\newcommand{\ttc}{\tau_{(1,j+1)}}

\newcommand{\FF}{\chi_{(1,j)}}
\newcommand {\aro}{(k)}
\newcommand {\da}{\partial}
\newcommand{\ppp}{\mbox{\hspace{5mm}}}
\newcommand{\ooo}{\mbox{\hspace{3mm}}}
\newcommand{\ooa}{\mbox{\hspace{1mm}}}

\section{Setting the problem for a completed scattering} \label{a1}

Let us consider a particle tunneling through the time-independent potential barrier
$V(x)$ confined to the finite spatial interval $[a,b]$ $(a>0)$; $d=b-a$ is the
barrier width. Let its in state, $\Psi_{in}(x),$ at $t=0$ be the normalized function
$\Psi^{(0)}_{left}(x)$ to belong to the set $S_{\infty}$ consisting from infinitely
differentiable functions vanishing exponentially in the limit $|x|\to \infty$. The
Fourier-transform of such functions are known to belong to the set $S_{\infty}$ as
well. In this case the position and momentum operators both are well-defined. Without
loss of generality we will suppose that
\begin{equation*}
\fl <\Psi^{(0)}_{left}|\hat{x}|\Psi^{(0)}_{left}>=0,\ppp
<\Psi^{(0)}_{left}|\hat{p}|\Psi^{(0)}_{left}> =\hbar k_0 > 0,\ppp
<\Psi^{(0)}_{left}|\hat{x}^2|\Psi^{(0)}_{left}> =l_0^2,
\end{equation*}
here $l_0$ is the wave-packet's half-width at $t=0$ ($l_0<<a$); $\hat{x}$ and
$\hat{p}$ are the operators of the particle's position and momentum, respectively.

Since we study a complete scattering, an important restriction should be imposed on
the rate of spreading the incident wave packet. Namely, we will suppose that the
average velocity $\hbar k_0/m$ is large enough, so that the parts of the incident wave
packet lying behind its CM, within the wave-packet's half-width, move toward the
barrier together with the CM; $m$ is the particle's mass.

As is known, the formal solution to the temporal one-dimensional Schr\"odinger
equation (OSE) of the problem can be written as $e^{-i\hat{H}t/\hbar}\Psi_{in}(x).$
In order to solve explicitly this equation we will use here the variant (see
\cite{Ch1}) of the well-known transfer matrix method \cite{Mez} that allows one to
calculate the tunneling parameters, as well as to connect the amplitudes of outgoing
and corresponding incoming waves, for any system of potential barriers.

Let $E$ be the energy of a particle. Then for the wave function $\Psi_{full}$ to
describe its stationary state in the out-of-barrier regions we have
\begin{eqnarray} \label{1}
\Psi_{full}(x;k)=\Api(k)e^{ikx}+\Amo(k)e^{-ikx}
\end{eqnarray}
for $x\le a$, and
\begin{eqnarray} \label{2}
\Psi_{full}(x;k)=\Apo(k)e^{ikx} +\Ami(k)e^{-ikx},
\end{eqnarray}
\noindent for $x>b$; here $k=\sqrt{2mE}/\hbar;$ $\Api(k)$ should be found from the
initial condition; $\Ami(k)=0.$ The coefficients entering this solution are connected
by the transfer matrix ${\bf Y}$:
\begin{eqnarray} \label{50}
\left(\begin{array}{c} \Api \\ \Amo
\end{array} \right)={\bf Y} \left(\begin{array}{c} \Apo \\ \Ami
\end{array} \right); \hspace{8mm}
{\bf Y}=\left(\begin{array}{cc} q & p \\ p^* & q^* \end{array} \right);
\end{eqnarray}
that can be expressed (see \cite{Ch1}) in terms of the real tunneling parameters $T$,
$J$ and $F$;
\newcommand{\iii}{\mbox{\hspace{10mm}}}
\begin{eqnarray} \label{500}
\fl q=\frac{1}{\sqrt{T(k)}}\exp\left[i(kd-J(k))\right];\ppp
p=\sqrt{\frac{R(k)}{T(k)}}\exp\left[i\left(\frac{\pi}{2}+ F(k)-ks\right)\right];
\end{eqnarray}
\noindent $T(k)$  (the real transmission coefficient) and $J(k)$ (phase) are even and
odd functions, respectively; $F(-k)=\pi-F(k)$; $R(k)=1-T(k)$; $s=a+b$. One can easily
show that for a particle impinging the barrier from the left
\begin{eqnarray} \label{700}
\Amo/\Api\equiv \amo= p^*/q, \ppp \Apo/\Api\equiv \apo= 1/q.
\end{eqnarray}
We will suppose that the tunneling parameters have already been calculated (in the
case of many-barrier structures, for this purpose one can use the recurrence relations
obtained in \cite{Ch1} just for these real parameters.

As is known, solving the TTP is reduced in the SWPA to timing a particle beyond the
scattering region where the exact solution of the OSE approaches the corresponding in
or out asymptote \cite{Tei}. Thus, definitions of characteristic times in this
approach can be done in terms of the in and out asymptotes of the tunneling problem.

Note, in asymptote in the one-dimensional scattering problem represents an one-packet
object to converge, well before the scattering event, with the incident wave packet.
But out asymptote represents the superposition of two non-overlapped wave packets to
converge, at $t\to \infty,$ with the transmitted and reflected ones. It is easy to
show that in asymptote $\Psi_{in}(x,t)$ and out asymptote $\Psi_{out}(x,t)$ can be
written, for the problem at hand, as follows
\begin{eqnarray} \label{59}
\fl \Psi_{in}(x,t)=\frac{1}{\sqrt{2\pi}}\int_{-\infty}^{\infty} f_{in}(k,t)
e^{ikx}dk,\ppp f_{in}(k,t)=\Api(k) \exp[-i E(k)t/\hbar];
\end{eqnarray}
\begin{eqnarray} \label{60}
\fl \Psi_{out}(x,t)=\frac{1}{\sqrt{2\pi}}\int_{-\infty}^{\infty} f_{out}(k,t)
e^{ikx}dk,\ppp f_{out}(k,t)= f_{out}^{tr}(k,t)+f_{out}^{ref}(k,t);
\end{eqnarray}
\begin{eqnarray} \label{61}
\fl f_{out}^{tr}(k,t)=\sqrt{T(k)}\Api(k) \exp[i(J(k)-kd-E(k)t/\hbar)];
\end{eqnarray}
\begin{eqnarray} \label{62}
\fl f_{out}^{ref}(k,t)=\sqrt{R(k)}\Api(-k)
\exp[-i(J(k)-F(k)-\frac{\pi}{2}+2ka+E(k)t/\hbar)]
\end{eqnarray}
where $E(k)=\hbar^2 k^2/2m.$

For a completed scattering we have
\[\fl \Psi_{full}(x,t)\approx\Psi_{in}(x,t)\ppp \mbox{when}\ppp t=0;\ppp
\Psi_{full}(x,t)=\Psi_{out}(x,t) \ppp\mbox{when}\ppp t\to\infty.\] It is obvious that
the larger is the distance $a$, the more correct is the approximation for
$\Psi_{full}(x,t)$ at $t=0$.

For the average particle's position, well before the scattering event, we have
\begin{eqnarray} \label{63}
<\hat{x}>_{in}=\frac{\hbar k_0}{m}t
\end{eqnarray} (hereinafter, for any Hermitian operator $\hat{Q}$
\[<\hat{Q}>_{in}=\frac{<f_{in}|\hat{Q}|f_{in}>}{<f_{in}|f_{in}>};\]
similar notations are used below for the transmitted and reflected wave packets). The
averaging separately over the transmitted and reflected wave packets yields
\begin{eqnarray} \label{64}
<\hat{x}>^{tr}_{out}=\frac{\hbar t}{m}<k>^{tr}_{out} -<J^\prime(k)>^{tr}_{out}+d;
\end{eqnarray}
\begin{eqnarray} \label{65}
<\hat{x}>^{ref}_{out}=\frac{\hbar t}{m}<k>^{ref}_{out}
+<J^\prime(k)-F^\prime(k)>^{ref}_{out}+2a
\end{eqnarray}
(hereinafter the prime denotes the derivative with respect to $k$). Exps. (\ref{63})
--- (\ref{65}) yield the basis for defining the asymptotic tunneling times in the
SWPA.

\section{Timing the particle's motion in the framework of the standard wave-packet
analysis} \label{a12}

For the following it is convenient to derive again the SWPA's tunneling times. Their
derivation is known to be based on the standard in quantum mechanics timing procedure
which is dictated by the correspondence principle. Namely, by the analogue with
classical mechanics where timing the particle's motion is reduced to the analysis of
the function $x(t)$ ($x$ is the particle's position, $t$ is time), in quantum
mechanics characteristic times for a particle should be derived from studying the
temporal dependence of the expectation value of the position of a particle (or, what
is equivalent, from studying the temporal behavior of the CM of a wave packet to
describe its state). Besides, quantum theory implies calculating the error of the
timing, which should be based on the analysis of the temporal dependence of the
mean-square deviation for the position operator.

The standard timing procedure is evident to imply that the average value of the
particle's position has its primary physical sense (as the most probable position of
a particle) at all stages of its motion. For instance, for a free particle whose
state is described by a Gaussian-like wave packet, this requirement is fulfilled and,
as a consequence, no problem arises in timing its motion. However, an essentially
different situation arises in the case of a tunneling particle. Now, following the CM
of a wave packet to describe the whole ensemble of tunneling particles becomes
meaningless at some stages of scattering. In particular, after the scattering event,
when we deal with two scattered (transmitted and reflected) wave packets, the
averaging over the whole ensemble of particles has no physical sense. By this reason
the above timing procedure cannot be applied in this case.

Of course, at late times one can attempt to define individual average positions of
transmitted and reflected particles. However, in timing, this implies a separate
description of both the subensembles at the first stage of scattering, what is widely
accepted to be impossible in conventional quantum mechanics. As a result, it is not
clear how to apply the above timing procedure both to the whole ensemble and to its
parts, transmitted and reflected particles. This question remains open in the SWPA.
In this section, following this approach, we will simply take the incident wave
packet as the counterpart to the transmitted (reflected) one at the initial stage of
scattering.

So, let $Z_1$ be the spatial point to lie at some distance $L_1$ ($L_1\gg l_0$ and
$a-L_1\gg l_0$) from the left boundary of the barrier, and $Z_2$ be the point to lie
at some distance $L_2$ ($L_2\gg l_0$) from its right boundary. Following \cite{Ha1},
let us define the difference between the times of arrival of the CMs of the incident
and transmitted packets at the points $Z_1$ and $Z_2$, respectively (this time will
be called below as the "transmission time"). Analogously, let the "reflection time"
be the difference between the times of arrival of the CMs of the incident and
reflected packets at the same point $Z_1$.

Thus, let $t_1$ and $t_2$ be such instants of time that
\begin{equation} \label{8}
<\hat{x}>_{in}(t_1)=a-L_1; \ppp <\hat{x}>^{tr}_{out}(t_2)=b+L_2.
\end{equation}
\noindent Then, considering (\ref{63}) and (\ref{64}), one can write the
"transmission time" $\Delta t_{tr}$ ($\Delta t_{tr} =t_2 -t_1$) for the given
interval in the form
\begin{eqnarray} \label{9}
\Delta t_{tr}=\frac{m}{\hbar}\Bigg[\frac{<J^\prime>_{out}^{tr} +L_2}
{<k>_{out}^{tr}}+\frac{L_1}{k_0}+a\left(\frac{1}{<k>_{out}^{tr}}
-\frac{1}{k_0}\right)\Bigg].
\end{eqnarray}
Similarly, for the reflected packet, let $t^{\prime}_1$ and $t^{\prime}_2$ be such
instants of time that
\begin{equation} \label{10}
<\hat{x}>_{in}(t^{\prime}_1) =<\hat{x}>_{out}^{ref}(t^{\prime}_2)=a-L_1.
\end{equation}
\noindent From equations (\ref{63}), (\ref{65}) and (\ref{10}) it follows that the
"reflection time" $\Delta t_{ref}$ ($\Delta t_{ref}=t^{\prime}_2-t^{\prime}_1$) can
be written as
\begin{eqnarray} \label{11}
\Delta t_{ref}=\frac{m}{\hbar}\Bigg[\frac{<J^\prime - F^\prime>_{out}^{ref} +L_1}
{<-k>_{out}^{ref}} +\frac{L_1}{k_0}+a\left(\frac{1}{<-k>_{out}^{ref}}
-\frac{1}{k_0}\right)\Bigg].
\end{eqnarray}

Note, the average values of $k$ for all three wave packets coincide only in the limit
$l_0\to\infty$ (i.e., for particles with a well-defined momentum). In the general
case these quantities are distinguished. For example, for a particle whose initial
state is described by the Gaussian wave packet, when
\[\Api(k)=A \exp(-l_0^2(k-k_0)^2), \ppp
A=\left(\frac{l_0^2}{\pi}\right)^{1/4},\] we have
\begin{equation} \label{100}
<k>_{tr}=k_0+\frac{<T^\prime>_{in}} {4l_0^2<T>_{in}};
\end{equation}
\begin{equation} \label{101}
<-k>_{ref}=k_0+\frac{<R^\prime>_{in}} {4l_0^2<R>_{in}}.
\end{equation}
Let
\[<k>_{tr}=k_0+(\Delta k)_{tr},\ppp <-k>_{ref}=k_0+(\Delta k)_{ref},\]
\noindent then relations (\ref{100}) and (\ref{101}) can be written in the form (note
that $R^\prime=-T^\prime$)
\begin{equation} \label{102}
<T>_{in}\cdot (\Delta k)_{tr}=-<R>_{in}\cdot(\Delta k)_{ref}
=\frac{<T^\prime>_{in}}{4l_0^2}.
\end{equation}

As is seen, quantities (\ref{9}) and (\ref{11}) cannot serve as characteristic times
for a particle. Due to the last terms in these expressions the above times depend on
the initial distance between the wave packet and barrier, with $L_1$ being fixed.
These terms are dominant for the sufficiently large distance $a$. Moreover, one of
them must be negative. For example, for the transmitted wave packet it takes place in
the case of the under-barrier tunneling through an opaque rectangular barrier, when
the difference $<k>_{out}^{tr}-k_0$ is sufficiently large. The numerical modeling of
tunneling \cite{Ha2,Ha1,Ter,Le1} shows in this case a premature appearance of the CM
of the transmitted packet behind the barrier, what just points to the lack of a
causal link between the transmitted and incident wave packets (see \cite{La1}).

As was shown in \cite{Ha2,Ha1}, this effect disappears in the limiting case
$l_0\to\infty$. For example, in the case of Gaussian wave packets the fact that the
last terms in (\ref{9}) and (\ref{11}) tend to zero when $l_0\to \infty$, with the
ratio $l_0/a$ being fixed, can be proved with help of Exps. (\ref{100}) and
(\ref{101}) (note that the limit $l_0\to \infty$ with a fixed value of $a$ is
unacceptable in this analysis, because it contradicts the initial condition $a\gg
l_0$ for a completed scattering).

However, one has to stress that in the limit $l_0\to\infty$ the incident and
transmitted (reflected) wave packets are causally disconnected, as in the general
case. As before, the former cannot be considered as the counterpart to the
transmitted (or reflected) wave packet at the initial stage of scattering. The
inconsistency of introducing the phase times in the SWPA is seen explicitly in the
particular case of the $\delta$-potential. As is known, the phase transmission time
is nonzero in this case. This result is usually treated as the evidence of a
non-local character of tunneling a particle. However, one has to bear in mind the
fact that in the SWPA this nonzero time describes the motion of the wave-packet's CM.
Thus, this result is clearly erroneous, since any point of a moving wave packet
should cross instantly the spatial region of a zero width.

At the same time we have to stress that, although Exps. (\ref{9}) and (\ref{11})
cannot be applied to a particle, they notwithstanding correctly describe the relative
motion of the transmitted (or reflected) and incident wave packets. Thus, the main
problem is that quantum mechanics, as it stands, does not provide a clear
prescription to interpret properly the behaviour of these packets as applied to a
particle. To clarify this question is the main goal of our study.

\section{A renewed wave-packet analysis}\label{a2}

\subsection{Tunneling a quantum particle as a combined process consisting from two
alternative elementary ones, transmission and reflection} \label{a21}

\subsubsection{Born's formula underlying the statistical interpretation of a wave
function and the problem of defining the expectation values of physical observables
for a tunneling particle.}

As is seen from the previous section, modeling the tunneling process in terms of wave
packets meets a paradoxical situation. On the one hand, since this process is
one-particle, the corresponding time-dependent state of a quantum particle is supposed
to be the quantum analogue of a classical one-particle trajectory. On the other hand,
the average values of the particle's position and momentum calculated for this
one-particle state, by Born's formula, cannot be interpreted as the {\it expectation}
values of these quantities.

However, as is well-known, this formula is an integral part of Born's statistical
interpretation of a wave function. Therefore the above controversy means that the
wave function to describe the one-particle tunneling process cannot be associated with
the motion of a single quantum particle. This is seen also from the fact that at late
times this function represents the superposition of the transmitted and reflected
wave packets. The naive statistical interpretation of this wave function (as that to
describe the motion of a single particle), which ignores the above controversy, leads
to the non-local pseudo-effect: transmitted and reflected particles of the ensemble
are inextricably intertwined even after the scattering event. Of course, the above
fact that Born's formula is not applicable to this case says that this "non-locality"
as well as the above interpretation of wave function have no physical sense.

The above situation is similar to that to arise in the well-known two-slits
experiment where the beam of mutually non-interacting, identically prepared particles
diffracts on an opaque screen with two slits. As is known, a wave packet to describe
this (one-particle) process passes, like usual waves, through both the slits. As in
the above case, the naive statistical interpretation of this wave function (as that to
describe the stochastic motion of a single particle) "permits" the particle to pass
simultaneously through both the slits; for both the events are inextricably
intertwined in this interpretation. That is, again we arrive at the pseudo-effect of
non-locality.

\subsubsection{Elementary and combined one-particle scattering states and processes.}

So, scattering a wave packet on the potential barrier as well as its scattering on
the slits in an opaque screen, each cannot be associated unambiguously with the
motion of a single quantum particle. This is true also for any one-particle quantum
scattering process which involves either several sinks or several sources of
particles, or several possible channels of motion. In all non-relativistic scattering
processes, a quantum particle is evident to be an indivisible object. And, though we
can say nothing, before measuring, about the position of the particle, we can say {\it
a priori} that this object can be emitted only by one source of particles, can move
only along one scattering channel and can be absorbed only by one sink of particles.
This property should be considered as a distinctive of the motion of a single quantum
particle. It is evident that the behaviour of wave packets in the case of the
tunneling process and above diffraction on slits does not guarantee the fulfillment
of this property.

By our approach, all quantum one-particle scattering processes and the corresponding
time-dependent states fall into two classes - elementary and combined. Namely, if the
above property is fulfilled, we deal with elementary ones. Otherwise, a quantum
one-particle scattering process and the corresponding state are combined. Unlike an
elementary process, a combined one implies several alternative scattering channels for
a particle. The motion of a particle along each channel should be described by the
corresponding wave function to obey the Schr\"odinger equation and the corresponding
boundary conditions to distinguish this channel from others. The wave function to
describe a combined process is the sum of those to describe all elementary ones
involved in the former.

Note, since all elementary processes involved in the same combined one are mutually
exclusive, a particle can take part only in one of them. This means that the whole
quantum ensemble of particles involved in the combined scattering process can be
uniquely decomposed into the subensembles of particles taking part only in one of the
elementary processes: the sum of norms of wave functions to describe the elementary
processes is equal to unit. The number of particles in each sumensemble is evident to
be constant in time.

We have to stress that only elementary one-particle states can be viewed as the
quantum analogue of classical ones. To calculate the expectation values of the
position and momentum of a particle is meaningful only for elementary processes. In
this case one can calculate the trajectory of the wave-packet's CM. It can be used
then for timing a quantum particle. As regards a combined process, making use of
Born's formula has no physical sense in this case. Such a process can be associated
with several CM's trajectories, rather than with one. The number of these
trajectories is equal to that of elementary processes involved in the combined one.

So, by our approach tunneling a non-relativistic quantum particle through an
one-dimensional static potential barrier is a combined process consisting from two
alternative elementary ones, transmission and reflection. Our next step is to find two
solutions to the Schr\"odinger equations to describe separately transmission and
reflection. The wave function for transmission will be named further as the
transmission wave function (TWF), and that for reflection will be named as the
reflection wave function (RWF). The main thing which should be taken into account in
finding these solutions is that the RWF should describe only reflected particles, and
the TWF does only transmitted particles. In both the cases, stationary solutions
should contain one incoming and one outgoing wave. In this paper we show that such
solutions do exist.

\subsection{Wave functions for one-dimensional transmission and
reflection}\label{a22}

So, let $\Psi_{tr}$ and $\Psi_{ref}$ be the searched-for TWF and RWF, respectively.
In line with subsection \ref{a21}, their sum represents the wave function to
describe, in the problem at hand, the combined state of the whole ensemble of
particles. Hence, from the mathematical point of view our task now is to find such
solutions $\Psi_{tr}$ and $\Psi_{ref}$ to the Schr\"odinger equation that for any $t$,
\begin{equation} \label{261}
\Psi_{full}(x,t)=\Psi_{tr}(x,t)+\Psi_{ref}(x,t)
\end{equation}
where $\Psi_{full}(x,t)$ is the full wave function to describe all particles (see
section \ref{a1}). In the limit $t\to \infty$
\begin{equation} \label{262}
\Psi_{tr}(x,t)=\Psi_{out}^{tr}(x,t); \ppp \Psi_{ref}(x,t)=\Psi_{out}^{ref}(x,t)
\end{equation}
where $\Psi_{out}^{tr}(x,t)$ and $\Psi_{out}^{ref}(x,t)$ are the transmitted and
reflected wave packets whose Fourier-transforms presented in (\ref{61}) and
(\ref{62}).

As is known, searching for the wave functions in the case of the time-independent
potential $V(x)$ is reduced to the solution of the corresponding stationary
Schr\"odinger equation. For a given $k$, let us find firstly the functions
$\Psi_{ref}(x;k)$ and $\Psi_{tr}(x;k)$ for the spatial region $x\le a$. In this
region let \begin{eqnarray} \label{265} \Psi_{ref}(x;k)=\Api\left(\Api^{ref}e^{ikx}
+\Amo^{ref}e^{-ikx}\right)
\end{eqnarray}
\begin{eqnarray} \label{2650}
\Psi_{tr}(x;k)=\Api\Big(\Api^{tr}e^{ikx}+ \Amo^{tr}e^{-ikx}\Big)
\end{eqnarray}
where $\Api^{tr}+\Api^{ref}=1$, $\Amo^{tr}+\Amo^{ref}=p^*/q$.

Since the RWF describes the state of reflected particles only, the probability flux
for $\Psi_{ref}(x;k)$ should be equal to zero, i.e.,
\begin{eqnarray} \label{264}
|\Api^{ref}|^2-|\Amo^{ref}|^2=0.
\end{eqnarray}
In its turn, for $\Psi_{tr}(x;k)$ we have
\begin{eqnarray} \label{263}
|\Api^{tr}|^2-|\Amo^{tr}|^2=\frac{\hbar k}{m}T(k)
\end{eqnarray}
(the probability flux for the full wave function $\Psi_{full}(x;k)$ and for
$\Psi_{tr}(x;k)$ should be the same).

Taking into account that $\Psi_{tr}=\Psi_{full}-\Psi_{ref}$ let us now exclude
$\Psi_{tr}$ from Eq. (\ref{263}). As a result, we obtain for $\Psi_{ref}$ the equation
\begin{eqnarray} \label{2630}
Re\left(\Api^{ref}-\Amo^{ref}\amo^* \right)=0.
\end{eqnarray}
Eq. (\ref{2630}) guarantees the coincidence of the probability flux for
$\Psi_{full}(x)$ and $\Psi_{tr}(x)$.

From condition (\ref{262}) for $\Psi_{ref}(x;k)$ it follows that
$\Amo^{ref}(k)=\amo(k)\equiv p^*/q$ (see (\ref{700})). Then Eq. (\ref{2630}) yields
that $Re(\Api^{ref})=R$, and Eq. (\ref{264}) leads to $|\Api^{ref}|^2=
|\Amo^{ref}|^2=|p^*/q|^2=R.$ Thus, $\Api^{ref}=\sqrt{R}(\sqrt{R}\pm i\sqrt{T}) \equiv
\sqrt{R}\exp(i\lambda)$; $\lambda=\pm\arctan(\sqrt{T/R})$.

As is seen, the superposition of the incoming waves to describe transmission and
reflection for a given $E$ yields the incoming wave of unite amplitude, which
describes the whole ensemble of incident particles. In this case, not only
$\Api^{tr}+\Api^{ref}=1$, but also $|\Api^{tr}|^2+|\Api^{ref}|^2=1$!

So, there are two solutions to satisfy the above requirements for $\Psi_{ref}(x;k),$
in the region $x\le a$. Considering Exps. (\ref{500}) for the elements $q$ and $p$,
we have
\begin{eqnarray} \label{266}
\Psi_{ref}(x;k)=-2\sqrt{R}\Api\sin\Bigg(k(x-a)+\frac{1}{2}
\left(\lambda-J+F-\frac{\pi}{2}\right)\Bigg) e^{i\phi_{(+)}}
\end{eqnarray}
where
\[\phi_{(\pm)}=\frac{1}{2}\left[\lambda \pm \left(J-F-\frac{\pi}{2}
+2ka\right) \right].\]

Now we have to show that only one of these solutions describes reflection. To select
it, we have to study both the solutions in the region $x\ge b$ where they can be
written in the form
\begin{eqnarray} \label{267}
\Psi_{ref}(x;k)=\Api\left(\Apo^{ref}e^{ikx}+ \Ami^{ref}e^{-ikx}\right);
\end{eqnarray}
\[\Apo^{ref}=\sqrt{R} G^* e^{i\phi_{(+)}}; \ppp
\Ami^{ref}=\sqrt{R} G e^{i\phi_{(+)}},\ppp G=q e^{-i\phi_{(-)}}-p^* e^{i\phi_{(-)}}.\]
Considering Exps. (\ref{500}) as well as the equality $\exp(i\lambda)=\sqrt{R}\pm
i\sqrt{T}$, one can show that
\[G=\mp i\exp\left[i\left(kb-\frac{1}{2} \left(J+F+\frac{\pi}{2}-\lambda
\right)\right) \right];\] here the signs ($\mp$) correspond to those in the
expression for $\lambda$.  Then, for $x\ge b$, we have
\begin{eqnarray} \label{268}
\Psi_{ref}(x;k)=\mp 2\sqrt{R}\Api\sin\Bigg[k(x-b)+\frac{1}{2}
\left(J+F+\frac{\pi}{2}-\lambda\right)\Bigg] e^{i\phi_{(+)}}.
\end{eqnarray}
For the following it is convenient to go over to the variable $x'$: $x=x_{mid}+x'$
where $x_{mid}=(a+b)/2.$ Then, for $x'\le -d/2$, we have
\[\fl \Psi_{ref}(x')=-2\sqrt{R}\Api\sin\Big[\frac{1}{2}\big(kd+\lambda
-J-\frac{\pi}{2}\big)+\frac{F}{2}+kx'\Big] e^{i\phi_{(+)}},\] for $x'\ge d/2$
---
\[\fl \Psi_{ref}(x')=\pm 2\sqrt{R}\Api\sin\Big[\frac{1}{2}\big(kd+ \lambda
-J-\frac{\pi}{2}\big)-\frac{F}{2}-kx'\Big] e^{i\phi_{(+)}}.\] From these expressions
it follows that for any point $x'=x_0$ ($x_0\le -d/2$) we have
\begin{eqnarray} \label{269}
\fl \Psi_{ref}(x_0)=-2\sqrt{R}\Api\sin\Big[\frac{1}{2}\big(kd+ \lambda
-J-\frac{\pi}{2}+F\big)+k x_0\Big] e^{i\phi_{(+)}}
\end{eqnarray}
\begin{eqnarray} \label{270}
\fl \Psi_{ref}(-x_0)=\pm 2\sqrt{R}\Api\sin\Big[\frac{1}{2}\big(kd+ \lambda
-J-\frac{\pi}{2}+F\big)+kx_0-F\Big] e^{i\phi_{(+)}}.
\end{eqnarray}

Let us consider the case of symmetric potential barriers: $V(x')=V(-x')$. For such
barriers the phase $F$ is equal to either 0 or $\pi$. Then, as is seen from Exps.
(\ref{269}) and (\ref{270}), one of the above two stationary solutions
$\Psi_{ref}(x';k)$ is odd in the out-of-barrier region, but another function is even.
Namely, when $F=0$ the upper sign in (\ref{270}) corresponds to the odd function, the
lower gives the even solution. On the contrary, when $F=\pi$ the second root
$\lambda$ leads to the odd function $\Psi_{ref}(x';k)$.

It is evident that in the case of symmetric barriers both the functions keep their
"out-of-barrier symmetry" in the barrier region as well. Thus, the odd solution
$\Psi_{ref}(x';k)$ is equal to zero at the point $x'=0$. Of importance is the fact
that this property takes place for all values of $k$. In this case the probability
flux, for any time-dependent wave function formed only from the odd (or even)
stationary solutions $\Psi_{ref}(x';k)$, should be equal to zero at the barrier's
midpoint. This means that particles impinging a symmetric barrier from the left are
reflected by the barrier without penetration into the region $x'\ge 0$. In its turn,
this means that the searched-for stationary-state RWF should be zero in the region
$x'\ge 0$, but in the region $x'\le 0$ it must be equal to the odd function
$\Psi_{ref}(x';k)$. In this case the corresponding probability density is everywhere
continuous, including the point $x'=0$, and the probability flux is everywhere equal
to zero.

As regards the searched-for TWF, $\Psi_{tr}(x;k)$, it can be found now from the
expression $\Psi_{tr}(x;k)=\Psi_{full}(x;k)-\Psi_{ref}(x;k)$. This function is
everywhere continuous, and the corresponding probability flux is everywhere constant
(we have to stress once more that this quantity has no discontinuity at the point
$x=x_{mid}$, though the first derivative of $\Psi_{tr}(x;k)$ on $x$ is discontinuous
at this point). Thus, as in the case of the RWF, wave packets formed from the
stationary-state TWF should evolve in time with a constant norm.

As is seen from Exps. (\ref{269}) and (\ref{270}), for asymmetric potential barriers,
both the solutions $\Psi_{ref}(x';k)$ are neither even nor odd functions.
Nevertheless, it is evident that for any given value of $k$ one of these solutions
has opposite signs at the barrier's boundaries. This means that, for any $k$, there
is at least one point in the barrier region, at which this function is equal to zero.
However, unlike the case of symmetric barriers, the location of such a point depends
on $k$. Therefore the behavior of the time-dependent RWF in the barrier region is more
complicated for asymmetric barriers. Now the most right turning point for reflected
particles lies, as in the case of symmetric barriers, in the barrier region, but this
point does not coincide in the general case with the midpoint of this region.

To illustrate the temporal behavior of the wave functions $\Psi_{full}$, $\Psi_{tr}$
and $\Psi_{ref}$ in the case of symmetric barriers, we have considered the case of the
rectangular barrier of height $V_0$. In this case, the stationary-state wave function
$\Psi_{ref}(x;k)$, for $a\le x\le x_{mid}$, reads as
\begin{eqnarray} \label{271}
\Psi_{ref}=2\sqrt{R}\Api
e^{i\phi_{(+)}}\big[\cos(ka+\phi_{(-)})\sinh(\kappa d/2)\nonumber\\
-\frac{k}{\kappa}\sin(ka+\phi_{(-)})\cosh(\kappa d/2)\big]\sinh(\kappa(x-x_{mid}))
\end{eqnarray} where
$\kappa=\sqrt{2m(V_0-E)}/\hbar$ ($E<V_0$); and
\begin{eqnarray} \label{272}
\Psi_{ref}=-2\sqrt{R}\Api e^{i\phi_{(+)}}\big[\cos(ka+\phi_{(-)})\sin(\kappa
d/2)\nonumber\\+\frac{k}{\kappa}\sin(ka+\phi_{(-)})\cos(\kappa
d/2)\big]\sin(\kappa(x-x_{mid}))
\end{eqnarray}
where $\kappa=\sqrt{2m(E-V_0)}/\hbar$ ($E\ge V_0$). In both cases
$\Psi_{ref}(x;k)\equiv 0$ for $x\ge x_{mid}.$

We have calculated the spatial dependence of the probability densities
$|\Psi_{full}(x,t)|^2$ (dashed line), $|\Psi_{tr}(x,t)|^2$ (open circles) and
$|\Psi_{ref}(x,t)|^2$ (solid line) for the rectangular barrier ($V_0=0.3 eV$, $a=500
nm$, $b=505 nm$) and well ($V_0=-0.3 eV$, $a=500 nm$, $b=505 nm$). Figures 1 ($t=0$),
2 ($t=0.4 ps$) and 3 ($t=0.42 ps$) display results for the barrier, and figures 4
($t=0$), 5 ($t=0.4 ps$) and 6 ($t=0.43 ps$) display results for the well. In both the
cases, the function $\Psi_{full}(x,0)$ represents the Gaussian wave packet with
$l_0=7.5 nm$; the average kinetic energy is equal to $0.25 eV,$ both for the barrier
and well. Besides, in both cases, the particle's mass is $0.067 m_e$ where $m_e$ is
the mass of an electron.

As is seen from figures 1 and 4, the average starting points for the RWF and TWF
differ from that for $\Psi_{full}$ (remind that the latter, unlike the former, cannot
be unambiguously interpreted as the expectation value of the particle's position).
The main peculiarity of the transmitting wave packet is that it is slightly
compressed in the region of the barrier, and stretched in the region of the well.
Figure 7 shows that, at the stage of the scattering event ($t=0.4 ps$; see also
figure 2), the probability to find a transmitting particle in the barrier region is
larger than in the neighborhood of the barrier. This means that in the momentum space
this packet becomes wider when the ensemble of particles enters the barrier region.
For the well (see figure 8) there is an opposite tendency. Note that for the barrier
$<T>_{in}\approx 0.149$. For the well $<T>_{in}\approx 0.863$.

We have to stress once more that, by our approach, the initial state of a tunneling
particle can be presented as the superposition of those for transmission and
reflection: $\Psi_{full}(x,0)=\Psi_{tr}(x,0)+\Psi_{ref}(x,0).$ One might say that it
is meaningless to define the initial state of transmitted particles, since it is
impossible to predict the future of a starting particle. However, the destination of
quantum theory is to predict the behavior of the quantum ensembles of identically
prepared particles, rather than that of one particle. Main thing is that
$\Psi_{tr}(x,0)$ is the only initial wave packet, for the given potential $V(x)$,
which evolves isometrically (causally) into the transmitted one, and
$\Psi_{ref}(x,0)$ is the only one which evolves isometrically into the reflected wave
packet. A more detailed discussion of some aspects of the separation of transmission
and reflection is given in the last section of the paper.

\section{Exact and asymptotic tunneling times for transmission and
reflection} \label{a3}
\subsection{Exact tunneling times} \label{a31}

\hspace*{\parindent} So, we have found two causally evolving wave packets to describe
the subensembles of transmitted and reflected particles at all stages of tunneling. It
is evident that the given formalism may serve as the basis to solve the tunneling time
problem, since now one can follow the CMs of the wave packets to describe separately
reflection and transmission, at all stages of its motion.

Let $t^{tr}_1$ and $t^{tr}_2$ be such moments of time that
\begin{equation} \label{80}
\frac{<\Psi_{tr}(x,t^{tr}_1)|\hat{x}|\Psi_{tr}(x,t^{tr}_1)>}
{<\Psi_{tr}(x,t^{tr}_1)|\Psi_{tr}(x,t^{tr}_1)>} =a-L_1;
\end{equation}
\begin{equation} \label{81}
\frac{<\Psi_{tr}(x,t^{tr}_2)|\hat{x}|\Psi_{tr}(x,t^{tr}_2)>}
{<\Psi_{tr}(x,t^{tr}_1)|\Psi_{tr}(x,t^{tr}_1)>} =b+L_2,
\end{equation}
\noindent where $\Psi_{tr}(x,t)$ describes transmission. Then, one can define the
transmission time $\Delta t_{tr}(L_1,L_2)$ as the difference $t^{tr}_2(L_2)-
t^{tr}_1(L_1)$ where $t^{tr}_1(L_1)$ is the smallest root of Eq. (\ref{80}), and
$t^{tr}_2(L_2)$ is the largest root of Eq.  (\ref{81}).

Similarly, for reflection, let $t$ be such that
\begin{equation} \label{110}
\frac{<\Psi_{ref}(x,t)|\hat{x}|\Psi_{ref}(x,t)>}
{<\Psi_{ref}(x,t)|\Psi_{ref}(x,t)>}=a-L_1,
\end{equation}
\noindent then the reflection time $\Delta t_{ref}(L_1)$ can be defined as $\Delta
t_{ref}(L_1)=t^{ref}_2-t^{ref}_1$ where $t^{ref}_1$ is the smallest root, and
$t^{ref}_2$ is the largest root of Eq. (\ref{110}) (of course, if they exist, for the
wave-packet's CM may do not enter the barrier region).

It is important to emphasize that, due to conserving the norms of $\Psi_{tr}(x,t)$
and $\Psi_{ref}(x,t)$ both the characteristic times are non-negative for any distances
$L_1$ and $L_2$. Both the definitions are valid when $L_1=0$ and $L_2=0$. In this
case the quantities $\Delta t_{tr}(0,0)$ and $\Delta t_{ref}(0)$ yield, respectively,
exact transmission and reflection times for the barrier region. Both the
characteristic times show, in fact, the time spent by the corresponding CM in the
barrier region. Of course, one has to bear in mind that in the case of reflection the
CM of the wave packet may turn back without entering the barrier region: in this case
$\Delta t_{ref}(0)=0$. Of course, if $L_1$ is larger than the wave-packet's width,
$\Delta t_{ref}(L_1)\neq 0$.

\newcommand {\uta} {\tau_{tr}}
\newcommand {\utb} {\tau_{ref}}

\subsection{Asymptotic tunneling times}

It is evident that in the general case the above average quantities can be calculated
only numerically. At the sane time, for sufficiently large values of $L_1$ and $L_2$,
one can obtain the tunneling times $\Delta t_{tr}(L_1,L_2)$ and $\Delta t_{ref}(L_1)$
in more explicit form. Indeed, in this case, instead of the exact subensemble's wave
functions, we can use the corresponding in asymptotes derived in $k$-representation.
Indeed, now the "full" in asymptote, like the corresponding out asymptote, represents
the sum of two wave packets:
\[f_{in}(k,t)=f_{in}^{tr}(k,t)+f_{in}^{ref}(k,t);\]
\begin{eqnarray} \label{75}
f^{tr}_{in}(k,t)=\sqrt{T}\Api\exp[i(\Lambda -\alpha\frac{\pi}{2} - E(k)t/\hbar)];
\end{eqnarray}
\begin{eqnarray} \label{76}
f^{ref}_{in}(k,t)=\sqrt{R}\Api\exp[i(\Lambda- E(k)t/\hbar)];
\end{eqnarray}
$\alpha=1$ if $\Lambda\ge 0$; otherwise $\alpha=-1.$ Here the function $\Lambda(k)$
coincides, for a given $k$, with one of the functions, $\lambda(k)$ or $-\lambda(k)$,
for which $\Psi_{ref}(x;k)$ is an odd function (see above). One can easily show that
for both the roots
\[|\Lambda^\prime(k)|=\frac{|T^\prime|}{2\sqrt{R T}}.\]

A simple analysis in the $k$-representation shows that well before the scattering
event the average kinetic energy of particles in both subensembles (with the average
wave numbers $<k>^{tr}_{in}$ and $<k>^{ref}_{in}$) is equal to that for large times:
\[<k>^{tr}_{in}=<k>^{tr}_{out}, \ppp <k>^{ref}_{in}=-<k>^{ref}_{out}.\]
Besides, at early times
\begin{eqnarray} \label{73}
<\hat{x}>^{tr}_{in}=\frac{\hbar t}{m}<k>^{tr}_{in} -<\Lambda^\prime(k)>^{tr}_{in};
\end{eqnarray}
\begin{eqnarray} \label{74}
<\hat{x}>^{ref}_{in}=\frac{\hbar t}{m}<k>^{ref}_{in} -<\Lambda^\prime(k)>^{ref}_{in}
\end{eqnarray}

As it follows from Exps. (\ref{73}) and (\ref{74}), the average starting points
$x_{start}^{tr}$ and $x_{start}^{ref}$, for the subensembles of transmitted and
reflected particles, respectively, differ from that for all particles:
\begin{eqnarray} \label{730}
x_{start}^{tr}=-<\Lambda^\prime>^{tr}_{in},\ooo x_{start}^{ref}=
-<\Lambda^\prime>^{ref}_{in}.
\end{eqnarray}
The implicit assumption made in the SWPA that incident, as well as transmitted and
reflected particles start, on the average, from the same point does not agree with
this result. Of great importance here is that $x_{start}^{tr}$ and $x_{start}^{ref}$
are the initial values of $<\hat{x}>^{tr}_{in}$ and $<\hat{x}>^{ref}_{in}$,
respectively, which have the status of expectation values of the particle's position.
They behave causally in time. As regards the average starting point for the whole
ensemble of particles, its coordinate is the initial value of $<\hat{x}>_{in}$ which
behaves non-causally; for this quantity is not an expectation value of the particle's
position (see also the last section of this paper). By this reason the (asymptotic)
phase times obtained in the SWPA should be considered as ill-defined quantities, for
any wave packets.

Let us take into account Exps. (\ref{73}), (\ref{74}) and again analyze the motion of
a particle in the above spatial interval covering the barrier region.  In particular,
let us calculate the transmission time, $\uta$, spent (on the average) by a particle
in the interval $[Z_1, Z_2]$. It is evident that the above equations for the arrival
times $t^{tr}_1$ and $t^{tr}_2$, which correspond to the extreme points $Z_1$ and
$Z_2$, respectively, read now as
%\begin{equation} \label{22}
\[<\hat{x}>^{tr}_{in}(t^{tr}_1)=a-L_1;
%\end{equation}
%\begin{equation} \label{220}
\ppp<\hat{x}>^{tr}_{out}(t^{tr}_2)=b+L_2.
\]
%\end{equation}

\noindent Considering (\ref{73}) and (\ref{64}), we obtain from here that now the
transmission time is
\begin{eqnarray} \label{23}
\fl \uta(L_1,L_2)\equiv t^{tr}_2-t^{tr}_1=\frac{m}{\hbar
<k>^{tr}_{in}}\left(<J^\prime>^{tr}_{out} -<\Lambda^\prime>^{tr}_{in} +L_1+L_2
\right).
\end{eqnarray}
Similarly, for the reflection time $\utb(L_1)$ ($\utb=t^{ref}_2-t^{ref}_1$), we have
%\begin{equation} \label{24}
\[<\hat{x}>^{ref}_{in}(t^{ref}_1)=a-L_1,
%\end{equation}
%\begin{equation} \label{240}
\ppp<\hat{x}>^{ref}_{out}(t^{ref}_2)=a-L_1.
%\end{equation}
\]
\noindent Considering (\ref{74}) and (\ref{65}), one can easily show that
\begin{eqnarray} \label{25}
\fl \utb(L_1)\equiv t^{ref}_2-t^{ref}_1=\frac{m}{\hbar <k>^{ref}_{in}}\left(<J^\prime
- F^\prime>^{ref}_{out}-<\Lambda^\prime>^{ref}_{in} +2L_1\right).
\end{eqnarray}

The inputs $\uta^{as}$ ($\uta^{as}=\uta(0,0)$) and $\utb^{as}$
($\uta^{as}=\uta(0,0)$) will be named below as the asymptotic transmission and
reflection times for the barrier region, respectively:
\begin{eqnarray} \label{230}
\uta^{as}=\frac{m}{\hbar <k>^{tr}_{in}}\Big(<J^\prime>^{tr}_{out}
-<\Lambda^\prime>^{tr}_{in}\Big),
\end{eqnarray}
\begin{equation} \label{250}
\utb^{as}=\frac{m}{\hbar <k>^{ref}_{in}}\left(<J^\prime -
F^\prime>^{ref}_{out}-<\Lambda^\prime>^{ref}_{in}\right)
\end{equation}
Here the word "asymptotic" points to the fact that these quantities were obtained
with making use of the corresponding in and out asymptotes. Unlike the exact tunneling
times the asymptotic times can be negative by value.

The corresponding lengths $d_{eff}^{tr}$ and $d_{eff}^{ref},$
\begin{eqnarray} \label{251}
\fl d_{eff}^{tr}=<J^\prime>^{tr}_{out} -<\Lambda^\prime>^{tr}_{in},\ppp
d_{eff}^{ref}=<J^\prime-F^\prime>^{ref}_{out} -<\Lambda^\prime>^{ref}_{in},
\end{eqnarray}
can be treated as the effective barrier's widths for transmission and reflection,
respectively.

\subsection{Average starting points and asymptotic tunneling times for
rectangular potential barriers and $\delta$-potentials} \label{a33}

Let us consider the case of a rectangular barrier (or well) of height $V_0$ and
obtain explicit expressions for $d_{eff}(k)$ (now, both for transmission and
reflection, $d_{eff}(k)=J^\prime(k) -\Lambda^\prime(k)$ since $F^\prime(k)\equiv 0$)
which can be treated as the effective width of the barrier for a particle with a
given $k$. Besides, we will obtain the corresponding expressions for the expectation
value, $x_{start}(k)$, of the staring point for this particle:
$x_{start}(k)=-\Lambda^\prime(k)$. It is evident that in terms of $d_{eff}$ the above
asymptotic times for a particle with the well-defined momentum $\hbar k_0$ read as
\[\uta^{as}=\utb^{as}=\frac{m d_{eff}(k_0)}{\hbar k_0}.\]

Using the expressions for the real tunneling parameters $J$ and $T$ (see
\cite{Ch1,Ch4}), one can show that, for the below-barrier case ($E\le V_0$),
\[d_{eff}(k)=\frac{4}{\kappa}
\frac{\left[k^2+\kappa_0^2\sinh^2\left(\kappa d/2\right)\right]
\left[\kappa_0^2\sinh(\kappa d)-k^2 \kappa d\right]} {4k^2\kappa^2+
\kappa_0^4\sinh^2(\kappa d)}\]
\[x_{start}(k)= -2\frac{\kappa_0^2}{\kappa}
\frac{(\kappa^2-k^2)\sinh(\kappa d)+k^2 \kappa d \cosh(\kappa d)} {4k^2\kappa^2+
\kappa_0^4\sinh^2(\kappa d)}\] where $\kappa=\sqrt{2m(V_0-E)/\hbar^2};$ for the
above-barrier case ($E\ge V_0)$ ---
\[d_{eff}(k)=\frac{4}{\kappa} \frac{\left[k^2-\beta
\kappa_0^2\sin^2\left(\kappa d/2\right)\right]\left[k^2 \kappa d-\beta
\kappa_0^2\sin(\kappa d)\right]} {4k^2\kappa^2+\kappa_0^4\sin^2(\kappa d)}\]
\[x_{start}(k)= -2\beta \frac{\kappa_0^2}{\kappa} \cdot
\frac{(\kappa^2+k^2)\sin(\kappa d)-k^2 \kappa d \cos(\kappa d)} {4k^2\kappa^2+
\kappa_0^4\sin^2(\kappa d)}\] where $\kappa=\sqrt{2m(E-V_0)/\hbar^2};$ $\beta=1$ if
$V_0>0$, otherwise, $\beta=-1$. In both the cases $\kappa_0=\sqrt{2m|V_0|/\hbar^2}$.

It is important to stress that $d_{eff}\to d$ and $x_{start}(k) \to 0$, in the limit
$k\to \infty$. This property guarantees that for infinitely narrow in $x$-space wave
packets the average starting points for both subensembles will coincide with that for
all particles.  Note, for wells, the values of $d_{eff}$ and, as a consequence, the
corresponding asymptotic tunneling times are negative, in the limit $k\to 0$, when
$\sin(\kappa_0 d)<0$.

Note that for sufficiently narrow barriers and wells, namely when $\kappa d\ll 1$, we
have $d_{eff}\approx d$. For the starting point we have
\[x_{start}(k)\approx -\frac{\kappa_0^2}{2 k^2} d, \ppp
x_{start}(k)\approx -\beta \frac{\kappa_0^2}{2 k^2} d,\] for $E\le V_0$ and $E\ge
V_0$, respectively.

For wide barriers and wells, when  $\kappa d\gg 1$, we have $d_{eff}\approx 2/\kappa$
and $x_{start}(k)\approx 0$, for $E\le V_0$; and
\[\fl d_{eff}(k)\approx 4k^2 d \cdot
\frac{k^2-\beta\kappa_0^2 \sin^2(\kappa d/2)}{4k^2\kappa^2+\kappa_0^4 \sin^2(\kappa
d)};\ppp x_{start}(k)\approx \frac{2\beta\kappa_0^2 k^2 d \cos(\kappa d)}{4 k^2
\kappa^2+ \kappa_0^4 \sin^2(\kappa d)},\] for $E\ge V_0$.

It is important that for the $\delta$-potential, $V(x)= W \delta(x-a),$ we have
$d_{eff}\equiv 0$. That is, like the dwell and Larmor times, $\uta^{as}=0$ in this
case. Thus, though the ensemble of identically prepared particles spends nonzero time
to pass through this potential, each quantum particle of the ensemble spends no time
in its barrier region. While the quantum ensemble of particles interacts with the
$\delta$-potential, there is a nonzero probability to find a particle near this
barrier.

Note, unlike the first derivative of $\Psi_{full}(x,t)$ with respect to $x$, that of
$\Psi_{tr}(x,t)$ has equal values in the limits $x\to a\pm 0$. The average force
calculated for a particle in the state $\Psi_{tr}(x,t)$ is zero, for this potential.
That is, $\Psi_{tr}(x,t)$ describes that part of the incident wave packet, which does
not experience the action of the $\delta$-potential. Transmitting particles start, on
the average, from the point $x_{start}(k),$
\[x_{start}(k)=-\frac{2m\hbar^2 W}{\hbar^4 k^2+m^2W^2},\]
and moves then freely.

\section{Discussion and conclusions}

\subsubsection*{The problem of introducing characteristic times for a tunneling particle as
that of calculating the expectation values of its position and momentum}

The main idea underlying this paper is that tunneling a particle through an
one-dimensional static potential barrier is a combined stochastic process consisting
from two alternative elementary ones - transmission and reflection. We showed that
the wave function to describe the tunneling process can be uniquely decomposed into
two solutions of the Schr\"odinger equation for the given potential, which describe
separately transmission and reflection. We found both the solutions in the case of
symmetric potential barriers and introduced, according to the standard timing
procedure, the average (exact and asymptotic) transmission and reflection times.

By our approach, in the most cases, quantum one-particle scattering processes are just
combined. Each of them represents a complex stochastic process consisting from several
alternative elementary ones. The decomposition of a combined process into elementary
ones can be performed uniquely. Accordingly, the wave function to describe the
combined process can be uniquely presented as a sum of those to describe all the
elementary ones.

The main peculiarity of combined states is that the averaging over such states of the
particle's position and momentum, with help of Born's formula intended for
calculating the expectation values of physical observables, does not give in reality
expectation values of these quantities. Both the average values behave non-causally
in time. Strictly speaking, in the case of combined states, the particle's position
and momentum (though their operators are Hermitian) lost their primary status of
physical observables! As a result, timing a particle in such states is meaningless
too. Only for elementary quantum processes and states Born's formula and the timing
procedure are valid. In other words, only for elementary states the particle's
position and momentum (and other physical quantities with linear Hermitian operators)
have their primary status of observables, and, as consequence, there is no problem to
time the motion of a particle being in such states.

\subsubsection*{About some aspects of a superposition of the probability fields}

As is shown, the peculiarity of the wave functions for transmission and reflection is
that each of them contains only one incident and only one scattered wave packets. At
the same time it is evident that if a particle was prepared in the combined state
$\Psi_{tr}(x,0)$, then this packet would be divided by the barrier into two parts. Of
course, in the case considered the initial combined state of a particle is
$\Psi_{full}(x,0)$ but not $\Psi_{tr}(x,0)$. Nevertheless, we have to clear up the
principal difference taking place between a wave function to describe a combined
process and those to describe elementary ones involved in the former.

Let, for the given potential, in addition to the problem at hand where the amplitudes
of incoming and outgoing waves are, respectively,
\begin{eqnarray} \label{800}
\api=1, \ppp \amo=\frac{p^*}{q}, \ppp \apo=\frac{1}{q}, \ppp \ami=0,
\end{eqnarray}
we have two auxiliary scattering problems with amplitudes
\begin{eqnarray} \label{801}
\bpi^{ref}=\frac{|p|^2}{|q|^2},\ppp\bmo^{ref}=\frac{p^*}{q},\ppp
%\nonumber\\
\bpo^{ref}=0,\ppp\bmi^{ref}=\frac{p^*}{|q|^2},
\end{eqnarray}
and
\begin{eqnarray} \label{802}
\bpi^{tr}=\frac{1}{|q|^2},\ppp\bmo^{tr}=0,\ppp
%\nonumber\\
\bpo^{tr}=\frac{1}{q},\ppp\bmi^{tr}=-\frac{p^*}{|q|^2}
\end{eqnarray}
(the transfer matrix (\ref{50}) is evident to be the same for all three problems).

Note that in the first auxiliary problem the only outgoing wave coincides with the
reflected wave arising in (\ref{800}). And, in the second one, the only outgoing wave
coincides with the transmitted wave in (\ref{800}). It is evident that the sum of
these two functions results just in that to describe the state of a particle in the
original tunneling problem.

As is seen, the main peculiarity of the superposition of these two probability fields
is that due to interference their incoming waves in the region $x>b$ fully annihilate
each other (note that in the corresponding reverse motion they are outgoing waves).
The corresponding flux of particles is reoriented into the region $x<a$. Thus, the
initial probability fields (\ref{801}) and (\ref{802}) associated with the
transmitted and reflected wave packets are radically modified under the
superposition. In this case, the wave packet connected causally to the transmitted
(reflected) one is just $\Psi_{tr}(x,t)$ ($\Psi_{ref}(x,t)$).

Thus, we see that the sum of wave functions (\ref{801}) and (\ref{802}) can be
presented as that of the stationary-states RWF and TWF. As a result of reorienting
the probability fields, the squared amplitude of the incoming wave (in the region
$x<a$) associated with reflection increases due to interference from the initial
value $|\bpi^{ref}|^2$ ($=R^2$) (see (\ref{801})) to $|\bpi^{ref}|^2+|\bmi^{ref}|^2$
($=R^2+TR=R)$ (in the RWF). In the case of transmission, the corresponding quantity
increases from the initial value $|\bpi^{tr}|^2$ ($=T^2$) (see (\ref{802})) to
$|\bpi^{tr}|^2+|\bmi^{tr}|^2$ ($=T^2+T R=T)$ (in the TWF).

As is seen, in contrast to probability fields (\ref{801}) and (\ref{802}),
$\Psi_{tr}(x,t)$ and $\Psi_{ref}(x,t)$ should be considered as an inseparable pair:
they cannot evolve separately. Of course, in this case one would doubt the reality of
these fields. Indeed, they can cannot be observed separately. And, besides, being
involved in the combined state, they cannot be directly observed, at early stages of
scattering, because of interference. However, for any combined process, namely the
interference between wave fields to describe elementary sub-processes provides all
needed information to justify their existence.

Indeed, let $|\Psi^{exp}_{full}(x,t)|^2$ be the result of measuring
$|\Psi_{full}(x,t)|^2$. Then, using the distributions $|\Psi_{tr}(x,t)|^2$ and
$|\Psi_{ref}(x,t)|^2$ calculated beforehand, we can extract, from the experiment
data, the difference
$|\Psi^{exp}_{full}(x,t)|^2-|\Psi_{tr}(x,t)|^2-|\Psi_{ref}(x,t)|^2$ to describe
interference between the wave fields $\Psi_{tr}(x,t)$ and $\Psi_{ref}(x,t)$. By our
approach, it must have two important properties: 1) the integral of this difference
over the region $(-\infty,\infty)$ must be zero; 2) this difference must be nonzero
only for the first stages of scattering, and only for $x\le x_{mid}$. The first
property means that the whole ensemble of particles, in this scattering problem, can
be indeed divided into two subensembles described by the distributions
$|\Psi_{tr}(x,t)|^2$ and $|\Psi_{ref}(x,t)|^2$. The second property means that one of
them is indeed connected causally to the transmitted wave packet, and another evolves
causally into the reflected one. This means, in turn, that the above decomposition is
unique.

Note, this property is inherent only to combined processes. The elementary states
$\Psi_{tr}(x,t)$ and $\Psi_{ref}(x,t)$, themselves represent decompositions into the
orthogonal stationary states. However, the latter cannot be treated as elementary
states. For the interference between them does not have the above two properties.
They cannot be separated, in principle.

So, the wave functions $\Psi_{tr}(x,t)$ and $\Psi_{ref}(x,t)$ describe two real
processes to proceed simultaneously. Taking into account the fact that a wave function
describes the beam (or, ensemble) of identically prepared particles, rather than a
single quantum particle, one can interpret the found decomposition of
$\Psi_{full}(x,t)$ as follows. Namely, $\Psi_{tr}(x,t)$ describes that part of the
beam of mutually non-interacted particles prepared in the state $\Psi_{full}(x,0)$,
which is transmitted through the barrier. Similarly, $\Psi_{ref}(x,t)$ does the
reflected part of this beam.

At early times these parts of the beam move in the same spatial region. At this stage
of scattering, the study of the motion of both parts is reduced to the analysis of
the interference between them. At all stages they evolve irrespective of each other,
not "seeing" their own counterparts; just $\Psi_{tr}(x,t)$ ($\Psi_{ref}(x,t)$) is
causally connected to the transmitted (reflected) wave packet considered separately,
but not $\Psi_{full}(x,t)$. As well as in the superposition of free moving wave
packets they do not destroy each other (after their meeting in some spatial region
they move unaltered), in the superposition of the modified wave fields
$\Psi_{tr}(x,t)$ and $\Psi_{ref}(x,t)$ they do not influence each other, too.

It is not surprising that particles of both parts start, on the average, from the
spatial points to differ from the average starting point calculated for the whole beam
of particles. Firstly, $<\hat{x}>_{in}\neq
<T><\hat{x}>^{tr}_{in}+<R><\hat{x}>^{ref}_{in}$ due to interference; here $<T>$ and
$<R>$ are the norms of $\Psi_{tr}(x,t)$ and $\Psi_{tr}(x,t)$, respectively. And, what
is more important, among these three average quantities, only $<\hat{x}>^{tr}_{in}$
and $<\hat{x}>^{tr}_{in}$ have the physical meaning of the expectation values of the
particle's position. The behaviour of $<\hat{x}>_{in}$, being averaged over two
alternative processes, is not causal. It cannot be interpreted as the expectation
value of the particle's position.

The main point of our research is that any combined state represents a superposition
of elementary states which are distinguishable. As a consequence, by our approach, the
experimental study of the probability density for a particle taking part in a
combined process means, in fact, the observation of the interference between the
elementary states involved in the combined process. However, maxima and minima of the
interference pattern behave non-causally. Or, more correctly, only when we know all
information about each elementary state, which just behaves causally, we can
unambiguously interpret the behaviour of the interference pattern.

All this takes place, in particular, in the case of tunneling. By this approach, the
non-causal behaviour of a tunneling wave packet (which have been pointed out by
\cite{La2}) is explained by the fact that tunneling is a combined process. An
exhaustive explanation of this quantum effect can be achieved only in the framework
of a separate description of transmission and reflection. For only these processes are
elementary, and, as a consequence, namely they (and the corresponding probability
densities) behave causally.

\subsubsection*{About the perspective of studying the temporal and other aspects of
quantum one-particle scattering processes}

A simple analysis shows that the definitions of the asymptotic tunneling times given
in our approach differ essentially from their analogs known in the literature. At this
point we have to note once more that a correct timing of transmitted and reflected
particles implies the availability of a complete information about these subensembles
of particles at all stages of scattering. Making use, in the alternative approaches,
of the incident wave packet as the counterpart to the transmitted (or reflected) wave
packet at the first stage of scattering is clearly an inconsistent step. For the
former does not connected causally to the transmitted (or reflected) one (see also
\cite{La2}). Just the wave functions for transmission and reflection found in our
approach provides all needed information. Thus, we think that our definitions of
tunneling times have a more solid basis.

Of course, a final decision in the long-lived controversy in solving the TTP should be
made by a reliable experiment. In this connection, we have to note that the main ideas
of such approaches as \cite{But}, \cite{Mu3} and others whose formalism involves the
peculiarities of the measurement process, may be very useful in the following study
of the tunneling and other scattering processes treated as combined ones. Indeed, the
fact that our definitions of the tunneling times do not coincide, for example, with
those obtained in \cite{But} and \cite{Mu3} does not mean at all that the main ideas
underling our and these approaches contradict each other. Rather they are mutually
complementary. We think that namely in combination all these ideas will be useful in
studying quantum scattering processes.

So, it would be very useful to define the Larmor time and time-of-arrival distribution
on the basis of wave functions for transmission and reflection. It is evident that
the influence of an external magnetic field (or absorbing potential) on transmitted
and reflected particles should be different. Hence, the study of the interference
between transmission and reflection, at the first stages of scattering, might permit
us to check both our idea of separating these elementary processes and ways
\cite{But,Mu3} of introducing characteristic times, which differ from the standard
timing procedure. For the first case, for this purpose, the same magnetic field (or
absorbing potential) might be localized in two equivalent spatial regions lying on
the same distance from the midpoint of a symmetric potential barrier. In this case,
the symmetry of the original potential remains unaltered, and there is no principal
problem to find the wave functions for transmission and reflection.

As regards further development of our approach, it can be applied, in principle, to
any potential localized in the finite spatial region. In one dimension, it is
applicable to the potential steps and asymmetric potential barriers. No principal
difficulties should arise also in separating transmission and reflection in the case
of quasi-one-dimensional structures, when the potential energy of a particle depends
only on one coordinate. As regards the scattering problem with two slits in the
opaque screen, it seems to involve four elementary processes, transmission and
reflection for the first and second slits. Besides, scattering a particle on a
point-like obstacle, with a spherically symmetrical potential, seems to involve two
alternative elementary processes. In this case there is a plain to separate both the
processes. This plain must be parallel to the vectors $\vec{p}_{0}$ and
$[\vec{r}_{0}\times\vec{p}_{0}]$ and pass through the obstacle; here $\vec{r}_{0}$
and $\vec{p}_{0}$ are the average position and momentum of a particle calculated for
its initial combined state.

Of course, in the general case the problem of decomposing some combined process into
alternative elementary ones may be technically complicated. This task should be
considered, in every case, separately.

\section*{Figure captions}
\begin{verbatim}
\Figure{\label{fig1}The $x$-dependence of $|\Psi_{full}(x,t)|^2$ (dashed line) which
represents the Gaussian wave packet with $l_0=7.5 nm$ and the average kinetic
particle's energy $0.25 eV$, as well as $|\Psi_{tr}(x,t)|^2$ (open circles) and
$|\Psi_{ref}(x,t)|^2$ (solid line) for the rectangular barrier ($V_0=0.3 eV$, $a=500
nm$, $b=505 nm$); $t=0$.}
\end{verbatim}

\begin{verbatim}
\Figure{\label{fig2}The same as in \ref{fig1}, but $t=0.4 ps$.}
\end{verbatim}

\begin{verbatim}
\Figure{\label{fig3}The same as in \ref{fig1}, but $t=0.42 ps$.}
\end{verbatim}

\begin{verbatim}
\Figure{\label{fig4}The $x$-dependence of $|\Psi_{full}(x,t)|^2$ (dashed line) which
represents the Gaussian wave packet with $l_0=7.5 nm$ and the average kinetic
particle's energy $0.25 eV$, as well as $|\Psi_{tr}(x,t)|^2$ (open circles) and
$|\Psi_{ref}(x,t)|^2$ (solid line) for the rectangular well ($V_0=-0.3 eV$, $a=500
nm$, $b=505 nm$); $t=0$.}
\end{verbatim}

\begin{verbatim}
\Figure{\label{fig5}The same as in \ref{fig4}, but $t=0.4 ps$.}
\end{verbatim}

\begin{verbatim}
\Figure{\label{fig6}The same as in \ref{fig4}, but $t=0.43 ps$.}
\end{verbatim}

\begin{verbatim}
\Figure{\label{fig7}The same functions for the barrier region; parameters are the
same as for \ref{fig2}.}
\end{verbatim}

\begin{verbatim}
\Figure{\label{fig8}The same functions for the barrier region; parameters are the
same as for \ref{fig5}.}
\end{verbatim}

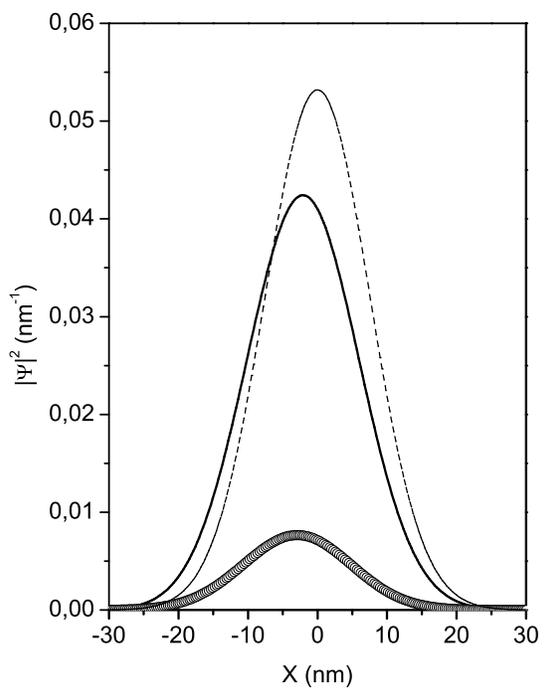
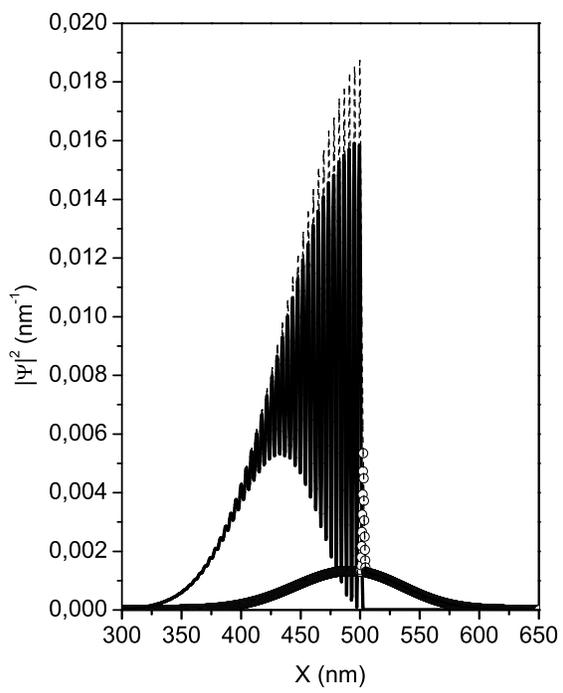
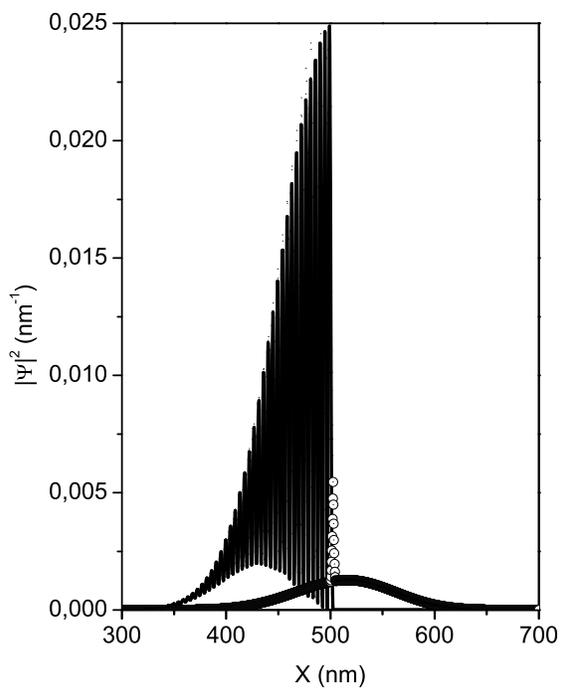
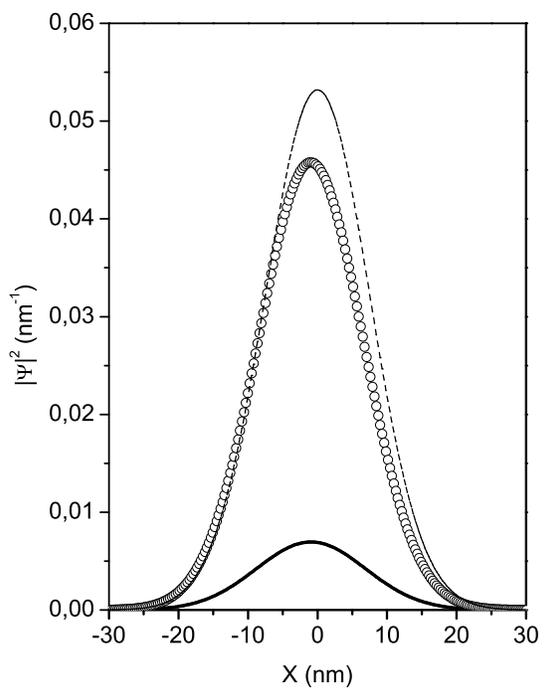
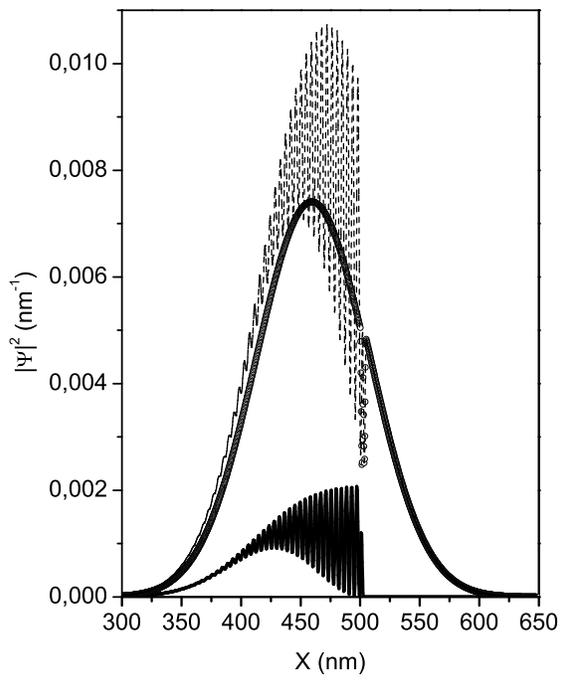
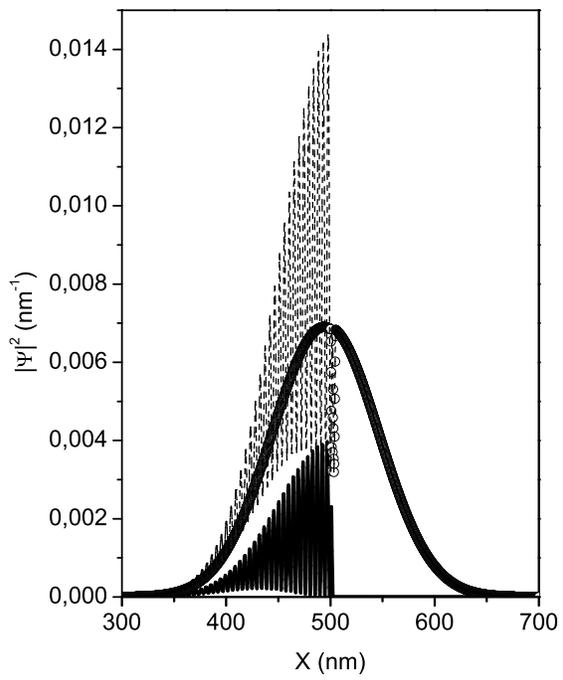
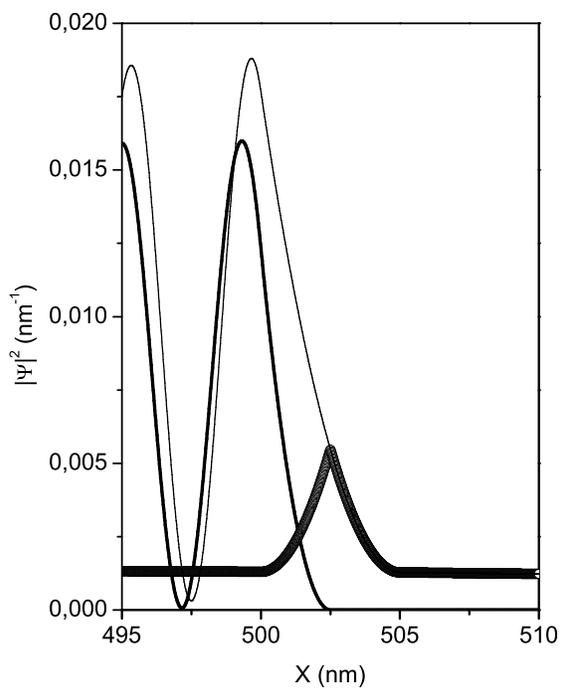
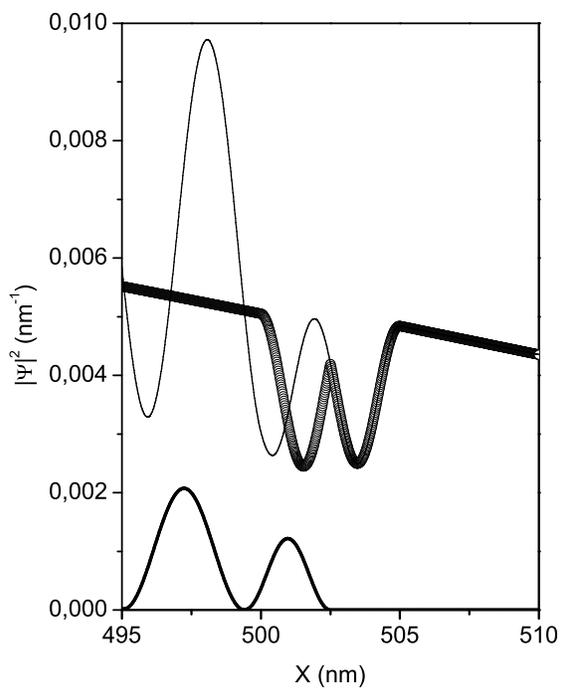
\end{document}